\documentclass[reprint, aps, prl,amsmath,amssymb, twocolumn, superscriptaddress]{revtex4-1}
\usepackage[utf8]{inputenc}
\usepackage[T1]{fontenc}
\usepackage{textcomp}
\usepackage{graphicx}
\usepackage{xcolor}
\usepackage{bbold}
\begin{document}

%\preprint{}

\title{Full control of polarization in tapered optical nanofibers}


\author{Maxime Joos}
%\thanks{}
\affiliation{Laboratoire Kastler Brossel, Sorbonne Université, CNRS, ENS-PSL Research University, Collège de France}
\author{Alberto Bramati}\affiliation{Laboratoire Kastler Brossel, Sorbonne Université, CNRS, ENS-PSL Research University, Collège de France}
\author{Quentin Glorieux}\affiliation{Laboratoire Kastler Brossel, Sorbonne Université, CNRS, ENS-PSL Research University, Collège de France}


\date{\today}

\begin{abstract}
We report on a protocol to achieve full control of the polarization in a nanofiber.
The protocol relies on studying the light scattered out from a nanofiber by means of two optical systems with $45\text{\textdegree}$ camera angle difference.
By studying the disturbance of the nanofiber on the radiation of embedded scatterers, we propose an explanation for the observed reduced scattering contrast of the nanofiber.
Thanks to this approach, we demonstrate an accuracy of the polarization control larger than 95\%.

\end{abstract}

\maketitle

\section{Introduction}
Optical nanofibers have drawn increasing attention in the last decade as a scalable light-matter interface.
They naturally provide large light-matter coupling strength due to the tight confinement of the guided light and enable easy interfacing with emitters through their evanescent fields \cite{SOLANO2017439, Nayak_2018}.
Recently, it was realized that optical nanofibers are also privileged interfaces to study and exploit light-matter chiral processes \cite{Petersen67, Sayrin, Volz, Lodahl2017}.
These developments require a precise control of the polarization of the light guided in the nanofiber. Primarily for the trapping geometry \cite{Kien2004PRA, Vetsch2010} and secondly for the probe beams in order to address specific radiative transitions.

Optical nanofibers are produced by locally heating and pulling commercial single-mode fibers \cite{Hoffman}. The guided light propagating towards or from the nanofiber is subject to birefringence \cite{Wolinski} of optical fibers.
Birefringence can be specific to the fiber, as for example the residual ellipticity of the core or internal stress.
Birefringence can else be due to the manipulation of the fiber (bending, twisting) or to its environment (temperature, pressure).
The polarization is therefore not maintained in general when light propagates through a fiber.
This issue makes it difficult to control the polarization in nanofibers: until now, the control has been limited to aligning the polarization quasi-linearly along two fixed orthogonal axis \cite{Vetsch2012, Goban}.
In this work, we show that the full control of the polarization can be achieved so that it is possible to deterministically set an arbitrary polarization in a nanofiber with good accuracy.

This new capability holds great promises in nanofiber experiments; it opens the way for exotic dipole trapping geometries \cite{REITZ2012}, magneto-optical trapping through the guided fields and enable to address new optical transitions.
Controlling the polarization in a nanofiber enables reciprocally to read an unknown polarization of the fundamental mode \cite{KIEN2004445} in a nanofiber.
Recently \cite{Joos2018}, this capability was used to analyze the emission properties of gold nanorods deposited at the surface of an optical nanofiber.

The paper is organized as follow: first, we formally introduce the strategy for compensating the birefringence in an optical fiber in order to control the polarization of the fundamental mode of a tapered optical nanofiber.
This procedure requires to characterize the birefringence of an optical system, which includes the optical fiber, and to correct it using a variable wave retarder (Berek compensator).

We explain the ideal procedure to do so by monitoring the scattering of the nanofiber region. 
Finally, we show how to adapt this procedure by taking into account the modified emission due to the nanofiber and  present compensation data with an accuracy better than 95\%. %à finir ...

\section{Matrix formulation of the problem}
\begin{figure}[htbp]
\def\svgwidth{\linewidth}
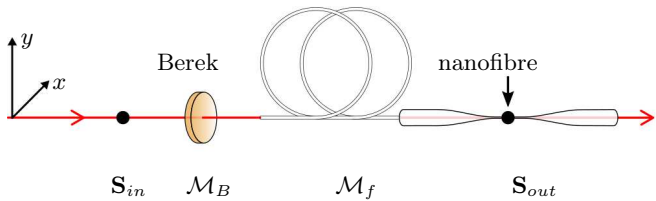
\caption{\textbf{Nanofiber setup.} Typical nanofiber setup to set the polarization of the guided light in the nanofiber region despite an arbitrary birefringence of the fiber. Control of the polarization before the fiber is achieved by usual means such as a combination of wave plates. 
The input polarization state is denoted $\mathbf{S}_{in}$. Light propagates through a Berek compensator (represented by the Müller matrix $\mathcal{M}_B$), a fiber system  (represented by $\mathcal{M}_f$) and ends up with polarization state $\mathbf{S}_{out}$ in the nanofiber region.}
\label{fig:1}
\end{figure}

Let consider the typical nanofiber setup presented in figure \ref{fig:1} where the light is coupled into a single mode fiber and guided trough the nanofiber.
Light also passes through several passive optical elements like mirrors or lenses, but these elements are not represented in figure \ref{fig:1}, for clarity.
The polarization of the free-space beam before the setup is controllable by usual means (for example using a combination of wave plates) and it is represented by the Stokes vector $\mathbf{S}_{in}$.
The polarization of the fundamental mode in the nanofiber is represented by the Stokes vector $\mathbf{S}_{out}$.
The general problem we discuss in this paper is: how to achieve a specific polarization of the guided light in the nanofiber region?

The strategy consists in transfering the controllable polarization state $\mathbf{S}_{in}$ in the nanofiber region.
This requires to compensate the birefringence introduced by the fiber system, represented by the Müller matrix $\mathcal{M}_f$.
The birefringence compensation is achieved by placing a Berek compensator --represented by the Müller matrix $\mathcal{M}_B$-- before the optical system.

The Berek compensator is a simple realization of a variable wave retarder.
It consists of an uni-axial crystal plate (typically calcite or magnesium fluoride) cut in such a way that its extraordinary axis is perpendicular to the optical faces.
Tilting the plate enables to introduce variable phase shift between the two polarization components while rotating the plate orient the optical axes.
In our experiment, we use a mechanical mount that enables to adjust each degree of freedom (retardance and optical axes orientation) independently (see for example the commercial model 5540M from Newport).
At our working wavelength ($\lambda = 638\text{ nm}$), the phase shift can be adjusted in the range $0-2\pi$.

In the following we recall the matrix formalism which describes the transformation of polarization by the system and detail a procedure to align the two degrees of freedom of the Berek compensator.
Within the (Stokes-Müller) matrix representation formalism, we can write: 
\begin{equation}
\mathbf{S}_{out}= \mathcal{M}_f \mathcal{M}_B \mathbf{S}_{in}.
\label{eq:1}
\end{equation}
The matrix $\mathcal{M}_f$ describes an arbitrary bent optical fiber which induces birefringence \cite{Ulrich:80} and optical rotation \cite{Ross1984}.
Birefringence will typically affect the ellipticity of the polarization while the optical rotation will rotates the polarization axes.
This is represented by the combination of a linear retarder and a rotator \cite{Ross1984, Jones1, Jones2}:
\begin{equation}
\mathcal{M}_f =\mathcal{S}(\theta) \mathcal{G}(\delta,\phi) \,
\label{eq:3}
\end{equation}
where $\mathcal{S}(\theta)$ is a rotation matrix:
\begin{equation}
\mathcal{S}(\theta) = \left(
\begin{array}{cccc}
1  & 0 & 0 & 0 \\
0  & \cos(2\theta)  & -\sin(2\theta) & 0\\
0  & \sin(2\theta) & \cos(2\theta) & 0 \\
0 & 0 & 0 & 1 \end{array}\\
\right),
\label{eq:2b}
\end{equation}
and $\mathcal{G}(\delta,\phi)$ describes a wave retarder with retardance $\delta$ and fast axis at angle $\phi$, and
\begin{equation}
\mathcal{G}(\delta,\phi) = \mathcal{S}(\phi)
 \left(
\begin{array}{cccc}
1  & 0 & 0 & 0 \\
0  & 1  & 0 & 0\\
0  & 0 & \cos(\delta) & -\sin(\delta) \\
0 & 0 & \sin(\delta) & \cos(\delta) \end{array}\\
\right)
 \mathcal{S}(-\phi).
\label{eq:2c}
\end{equation}
Hence the effect of the bent fiber on the polarization is determined by three parameters: the phase shift $\delta$ and the fast axis orientation $\phi$ of the equivalent retarder, and the rotation $\theta$ of the equivalent rotator.

The goal of the compensation procedure is to adjust the two degrees of freedom of the Berek compensator to induce a birefringence opposite to the equivalent retarder represented by $\mathcal{G}(\delta,\phi)$, i.e if $\mathcal{M}_B = \mathcal{G}(\delta,\phi)^{-1}$.
By doing so, the overall optical system of equation (\ref{eq:1}) reduces simply to an optical rotator:
\begin{equation}
\begin{aligned}
\mathcal{M}_f \mathcal{M}_B &= \mathcal{M}_f \mathcal{G}(\delta,\phi)^{-1}  = \mathcal{S}(\theta).
\label{eq:6}
\end{aligned}
\end{equation}
The polarization state in the nanofiber $\mathbf{S}_{out}$ is hence a rotated replica of the input polarization $\mathbf{S}_{in}$.
This rotation by an angle $\theta$ can be easily measured, and this third degree of freedom can be compensated using a single half-wave plate.
In this sense, we say to have the full control of the polarization state inside the nanofiber region.

\section{Berek Compensation}
The Berek compensation relies on studying the light scattered in the nanofiber region.
The microscopic origin of this scattering is not a well documented topic.
Surface imperfections and localized random inhomogeneities in the bulk were suggested to be at the origin of the scattering \cite{Vetsch2012}.
The presence of dopants and impurities cannot be excluded neither.
For now, we assume that scattering originates from point-like sources that conserve the polarization of the incident guided light.

\begin{figure}[htbp]
\def\svgwidth{\linewidth}
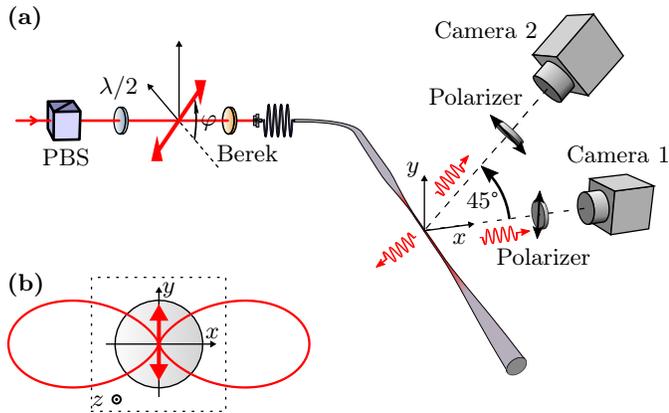
\caption{\textbf{Berek compensation setup.} \textbf{(a)} Experimental setup. Linearly polarized light is prepared with a polarized beam splitter PBS, followed by a half-wave plate and sent in the optical system including the Berek compensator and the fiber. 
A small fraction of the guided light in the nanofiber is scattered out and collected by two optical systems with different angles.
The cameras are equipped with polarizers filtering out the longitudinal (along $z$) component of the scattered light. 
Because scattering is supposed to be polarization maintaining, the study of the scattered light enables to infer the guided polarization. 
A position of the Berek compensator can be found to have the output polarization state $\mathbf{S}_{out}$ being a replica of the input polarization $\mathbf{S}_{in}$. \textbf{(b)} Transverse representation of the nanofiber showing the scattering pattern of the guided light in the case of vertical polarization of the fundamental mode (neglecting the longitudinal component of the field). We assume a dipole like radiation pattern, i.e. no alteration of the scattering by the nanofiber.}
\label{fig:2}
\end{figure}

Therefore, we can have access to some crucial information on the guided polarization by looking at the scattered light.
We further assume that the nanofiber does not alter the radiation pattern of the scatterers.
For example, if the guided light is vertically polarized in the nanofiber, as illustrated in figure \ref{fig:2}.(b), we assume the azimuthal radiation pattern to be dipole-like (following a pattern in $\cos^2$).

Consider now the fiber setup of figure \ref{fig:2}.(a) where two cameras are used to collect the scattered light from the nanofiber.
One camera is positioned at horizontal and faces the $x$ axis.
The second camera has a different angle and is positioned $45\text{\textdegree}$ above the first camera.

Both cameras look at the same region of the nanofiber near the origin  and have optical axes orthogonal to the nanofiber axis ($z$).
The cameras are further equipped with linear polarizers aligned perpendicularly with respect to the nanofiber $z\text{-axis}$ and to the optical axis of the camera.
In this position, the polarizers filter out the longitudinal (along $z$) component and transmit only the transverse component of the scattered light.
This filtering stage considerably simplify the analysis \cite{Vetsch2012}.

If we assume, as illustrated in figure \ref{fig:2}.(b), that the nanofiber does not alter the linear dipole radiation, the radiation patterns has a $\cos^2(\Theta)$ dependence.
Now, this is the same dependence as the projection of the dipole on the vertical axis.
Hence, the signal collected by the horizontal camera, called $I_1$, is proportional to the intensity of guided light polarized vertically. 
The signal collected by the second camera at $45\text{\textdegree}$, called $I_2$, is proportional to the intensity of light polarized at $135\text{\textdegree}$.
These two quantities are closely related to the normalized Stokes parameter $S_{out}^1$ and $S_{out}^2$ of the guided light.
From the definition of the Stokes parameters, one gets: 
\begin{equation}
    \begin{aligned}
        S_{out}^1 &= -\frac{I_1-\overline{I_1}}{\overline{I_1}}\\
        S_{out}^2 &= -\frac{I_2-\overline{I_2}}{\overline{I_2}}\\
    \end{aligned}
    \label{eq:8}
\end{equation}
where $\overline{I_1}$ and $\overline{I_2}$ are half of the maximum signals corresponding to vertically polarized light and linearly polarized light at $135\text{\textdegree}$ respectively.
Linear polarization of the guided light is characterized by having $(S_{out}^1)^2 + (S_{out}^2)^2 =1$.
This is the criteria, we will use to identify a linear polarization of the guided light.\\

Let us now consider the Berek compensator in the neutral position, equivalent to the situation in which the Berek is removed, and search for the linear input polarization angle, denoted $\varphi_0$, that gives rise to linear polarization in the nanofiber region.
In practice, linear polarization at angle $\varphi$ is prepared using a horizontal polarizer followed by a motorized half-wave plate (see figure \ref{fig:2}.(a)).
When, we observe a linear polarization in the nanofiber region, we know that the linear input polarization is aligned with respect to the optical axes of the effective wave retarder represented by $\mathcal{G}(\delta,\phi)$, i.e. $\varphi_0 = \phi$.

From this position, we rotate the input polarization by $45\text{\textdegree}$ which will, in general, gives rise to an elliptical polarization in the nanofiber because the fiber system acts as a variable-wave retarder (plus optical rotation).
For full birefringence compensation we must align the optical axis of the Berek with respect to $\varphi_0 = \phi$ and tilt the plate until we recover a linear polarization in the nanofiber region.
By doing so, we have introduced a retardance $-\delta$ opposite to the retardance of the fiber system $\mathcal{M}_f$.

This deductive procedure relies, for the derivation of equation (\ref{eq:8}), on the assumption that scatterers in the nanofiber have dipole-like radiation patterns (see figure \ref{fig:2}.(b)). 
As we will show in the following, the nanofiber significantly alters the radiation pattern of the embedded scatterers.
This effect needs to be taken into account and requires to adapt the Berek procedure presented above.

\section{Radiation pattern of dipoles in a nanofiber}
In order to have a better understanding on how the scattered light by the nanofiber gives information on the polarization of the guided light, we investigate numerically the radiation of individual linear dipoles located in the bulk of a nanofiber.
We then calculate and measure the radiation of a portion of the nanofiber, i.e. the average emission of a large number of  scatterers, randomly distributed across the nanofiber and sum along the profile imaged by an optical system.

\begin{figure}[htbp]
\scriptsize
\def\svgwidth{\linewidth}
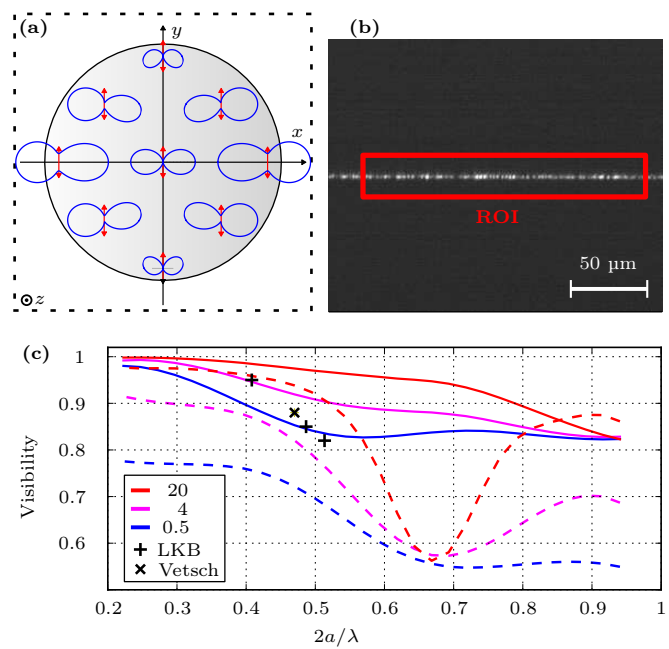
\caption{\textbf{Guided light scattering in a nanofiber}. \textbf{(a)} Far-field radiation patterns of linear dipoles located in the bulk of a nanofiber computed according to Mie theory. The radiation significantly differ from the free-space case and depends on the position of the scatterer in the transverse plane of the nanofiber. \textbf{(b)} Typical microscope image of a nanofiber scattering its guided light as seen from one of the cameras. The measured scattered signal correspond to the signal integrated over a region of interest (ROI) which includes the portion of nanofiber under investigation. \textbf{(c)} Visibility simulation of the radiation pattern of a portion of nanofiber scattering its guided light. We plot the visibility as a function of the nanofiber diameter, expressed in unit of $\lambda$. We consider different linear density of scatterer along the fiber expressed in terms of average number of scatterer per slice of nanofiber with thickness equal to the optical resolution. We consider scatterers randomly distributed: first, in the bulk (solid line) and second, on the surface (dashed line). We also plot measured visibilities observed in our lab (LKB) and in the literature (Vetsch, \cite{Vetsch2012}).}
\label{fig:3}
\end{figure}

To compute the dipole radiation patterns, we employ the diffraction theory of a plane wave by a dielectric cylinder, which we refer to as Mie theory \cite{Barber, Loo2018}.
We present in the figure \ref{fig:3}.(a) the far-field intensity radiation patterns of individual vertical dipoles in a nanofiber.
We assumed a silica nanofiber (refractive index $n=1.46$) with diameter of $300\text{ nm}$ surrounded by air or vacuum (refractive index $n=1.00$).
The scattered light is assumed to have a wavelength in vacuum of $638\text{ nm}$.
We consider dipoles at different location in the nanofiber section.
Only dipoles located exactly on the $y$ axis do not radiate light in the vertical direction (i.e. along the dipole orientation).
On contrary, dipoles not in the nanofiber center radiate along the dipole direction in strong contrast to the case of free space radiation.

We investigate how the individual scatterers, with unknown distribution within the fiber, contribute to the average signal recorded by the camera.
A typical microscope image of a nanofiber is presented in figure \ref{fig:3}.(b).
The scattering of light is not homogeneous over the length of the nanofiber which is consistent with the hypothesis of individual scatterers distributed randomly along the nanofiber.
To extract from the nanofiber image a useful signal, i.e. a signal that is not sensitive to the microscopic irregularity of the scattering, we have to integrate the image over a region of interest that includes a large number of scatterers (see figure \ref{fig:3}.(b)).
This portion of the nanofiber is typically a few hundred microns long.

We model the nanofiber portion as composed of a large number of adjacent cylindrical slices with thicknesses equal to the optical resolution of the optical system.
If there is no more than one scatterer per such slice, i.e if all the scatterers of the nanofiber are optically resolved, the scatterers contribute independently to the integrated camera signal.
In this case, we can calculate the integrated signal by summing the intensity radiation pattern of each individual scatterer, assuming that scatterers are randomly distributed in the transverse plane ($xy$) of the nanofiber.

In the case of scatterers that are not optically resolved, i.e if at least two scatterers are confined within such a slice of the nanofiber portion, these scatterers interfere on the image plane and contribute coherently to the detected signal.
The integrated camera signal would differ significantly from the previous case in a way that we will precise in the following.

A crucial parameter to characterize the radiation scattered by a portion of nanofiber is the visibility:
\begin{equation}
    V = \frac{I_{max}-I_{min}}{I_{max}+I_{min}}
\end{equation}
where $I_{max}$ is the maximum camera signal obtained when the linear dipoles are orthogonal to the camera optical axis.
$I_{min}$ is the minimum camera signal obtained when the linear dipoles point towards the camera.
In the limit of a nanofiber that has no effect on the radiation pattern of embedded linear dipoles, a visibility of 1 is expected, expressing the fact that there is no scattering in the direction of the linear dipole axes.

Based on Mie theory, we present in figure \ref{fig:3}.(c) simulations of the average visibility of a portion of nanofiber as a function of the nanofiber diameter (expressed in unit of $\lambda = \text{638 nm}$).
We consider various possible distributions for the scatterers.

Regarding the transverse distribution, we assume scatterers randomly and homegeneously distributed in the bulk of the nanofiber (solid lines of figure \ref{fig:3}.(c)) and scatterers randomly distributed at the surface of the nanofiber (dashed lines of figure \ref{fig:3}.(c)).
Considering the longitudinal distribution, we assume different densities: 0.5, 4 and 20 scatterers per unit of optical resolution.

Finally, we compare our simulations with the measured visibilities observed in our experiment (LKB) and in the literature by Vetsch et al. \cite{Vetsch2012} (Vetsch).
The measured visibilities are larger for thinner nanofibers which is a feature predicted by our simulations.

At low scatterer density, the general trend of our simulations is that the visibility drops with the diameter of the nanofiber.
For very small diameter ($2a/\lambda \ll 1$), the visibility tends to equal 1 meaning that the fiber does less alter the radiation of embedded scatterer.
A larger density of scatterer per unit of optical resolution increases also the visibility.
This is due to destructive interference in the far-field and in the direction of the dipoles as explained in the Annexe \ref{appendix:A}.

The measured visibilities are compatible with our simulations and support our model of light scattering by a nanofiber. 

This study enables to link the far-field radiation pattern to hypothesis on the microscopic properties of nanofibers, such as the distribution of the scatterer in the bulk of the nanofiber.

\section{Test of the compensation procedure}
As we saw above, the nanofiber is expected to significantly alter the radiation of the embedded scatterers.
Now, the Berek procedure described earlier relies on equation (\ref{eq:8}) that gives the Stokes parameters $S_{out}^1$ and $S_{out}^2$, which itself derive from the dipole-like radiation of the nanofiber.
Therefore, the Berek procedure is not accurate and needs to be adapted .
In the following, we present the experimental compensation procedure we applied and quantify its accuracy.

We prepare linear polarization with angle $\varphi$ at the input of the optical system and record the signals $I_1(\varphi)$ and $I_2(\varphi)$ of the cameras.
This is performed by a motorized rotation of a half-wave plate at the input of the system as presented in figure \ref{fig:2}.
Measured signals $I_1(\varphi)$ and $I_2(\varphi)$ as a function of the input polarization angle $\varphi$ are presented in figure \ref{fig:3}.
The measured signals present oscillations with a period of $180\text{\textdegree}$ from which we can compute the visibility:
\begin{equation}
    V_{1,2} = \frac{I_{1,2}^{max}-I_{1,2}^{min}}{I_{1,2}^{max}+I_{1,2}^{min}},
\end{equation}
where $I^{max}_{1,2}$ and $I^{min}_{1,2}$ are the maximum and minimum value of the signals collected on camera 1 and 2 respectively.

If the Berek compensator cancels properly the birefringence of the fiber system, linear polarizations in the input are transformed in linear polarizations in the nanofiber.
This situation corresponds to having maximal visibilities of the camera signals.
The Berek procedure can hence be simplified by searching to maximize the visibilities of the camera signals while rotating the input polarization.

\begin{figure}[htbp]
\tiny
\def\svgwidth{\linewidth}
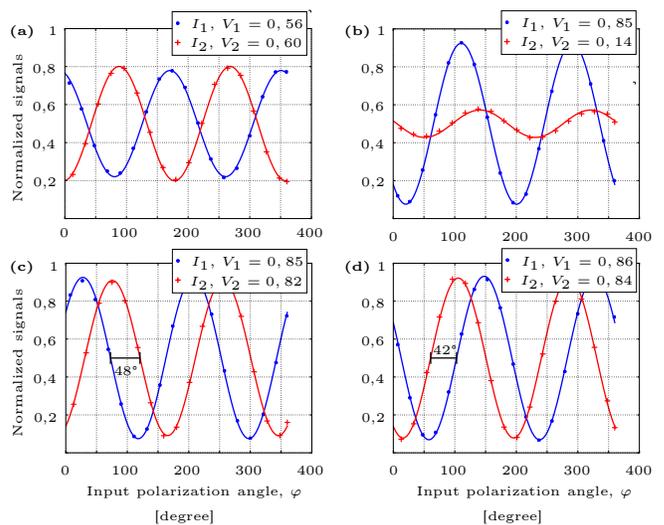
\caption{\textbf{Scattered signals and guided polarization}. Measured signals on camera 1 and camera 2 as the input linear polarization in rotated in the clockwise direction by an angle $\varphi$.\textbf{(a)} Visibilities are not maximal, which corresponds to a Berek adjustments that does not compensate the birefringence of the fiber. \textbf{(b)} Visibility is maximal on camera 1, not on camera 2. This adjustment of the Berek compensator enables to align the polarization along the $y$ ($I_1$ is maximal) or $x$ ($I_1$ is minimal) axis. \textbf{(c,d)} Visibilities are maximal on both cameras simultaneously, meaning that the polarization is linear in the nanofiber region. In \textbf{(c)}, the polarization in the nanofiber rotates in the counter-clockwise direction while the input polarization rotates in the clockwise direction, corresponding to the effect of a half-wave plate. In \textbf{(d)}, the polarization in the nanofiber rotates in the same direction as the input polarization, corresponding to a good adjustment of the Berek compensator.}
\label{fig:4}
\end{figure}

Figure \ref{fig:4}.(a) shows camera signals when the Berek compensator adjustment is not optimal: in the case where the fiber system is not compensated, elliptical polarization is obtained in the nanofiber region leading to visibilities which are not maximal.

Figure \ref{fig:4}.(b) correspond to an adjustment of the Berek compensator that maximize the visibility on camera 1 at horizontal.
Hence, this Berek adjustment enables to have the polarization of the fundamental mode aligned at the horizontal or at the vertical for the appropriate input linear polarization.

Figure \ref{fig:4}.(c-d) represent camera signals with maximum visibilities.

Note that, if the overall optical system (Berek + fiber system) introduces a half-wave retardation, we also expect linear polarization in the nanofiber region and hence, we will also observe maximal visibilities of the camera signals.

It is then crucial to be able to discriminate between this two adjustments of the Berek.
When the optical system acts as a half-wave plate, rotation of the input linear polarization leads to a reversed rotation of the linear polarization in the nanofiber.
Whereas when the optical system does not introduce any birefringence, i.e. is compensated, the rotation of the output and input linear polarizations occur in the same sense.

Since the sense of rotation of the input linear polarization is known (we set it to be clockwise), it is sufficient to observe the sense of rotation of the linear polarization in the nanofiber to be able to discriminate between the two cases in question.
The information on the sense of rotation of the linear polarization in the nanofiber is contained in the relative delay between the camera signals $I_1(\varphi)$ and $I_2(\varphi)$ as we will clarify now.

When linear polarization rotates in the nanofiber, both cameras record the same signal only phase shifted according to the difference of camera angle, which is, in our case $45\text{\textdegree}$.
If the signal $I_1(\varphi)$ is ahead by $45\text{\textdegree}$ with respect $I_2(\varphi)$, as illustrated in figure \ref{fig:4}.(c), it means that the polarization in the nanofiber rotates anti-clockwise.
If the signal $I_2(\varphi)$ is ahead by $45\text{\textdegree}$ with respect $I_1(\varphi)$, as illustrated in figure \ref{fig:4}.(d), the linear polarization in the nanofiber rotates anti-clockwise.
Experimentally, we observe 

The desired polarization mapping is hence ensured by two conditions:
\begin{itemize}
    \item The visibilities $V_{1,2}$ are maximal.
    \item $I_1(\varphi)$ is delayed from $I_2(\varphi)$ by $45\text{\textdegree}$.
\end{itemize}

The compensation procedure consists in fulfilling simultaneously these two conditions by playing on the two degrees of freedom of the Berek compensator.
The procedure is therefore not a deductive procedure anymore but rather an iterative process of convergence toward the best compensation.
Experimentally, we start by searching the Berek adjustment that ensures the correct delay between the camera signal $I_1(\varphi)$ and $I_2(\varphi)$ without considering the visibility yet.
We then focus on maximizing the visibilities while keeping the proper delay.

In the following, we experimentally demonstrate that the birefringence compensation procedure allows for correctly mapping arbitrary polarization states from and to the nanofiber. 
We also evaluate the typical accuracy of the mapping.

\begin{figure}[htbp]
% \centering
\small
\def\svgwidth{\linewidth}
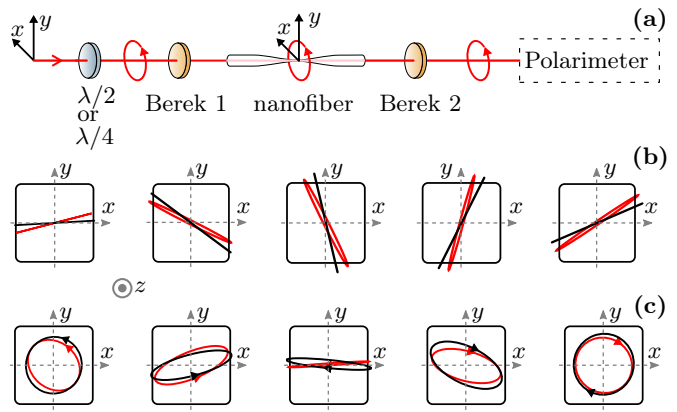
\caption{\textbf{Test of the birefringence compensation}. \textbf{(a)} Experimental setup to test birefringence compensation. Each portion of the fiber setup (on the left and right side of the nanofiber) is compensated independently using Berek compensator 1 and 2 respectively. The goal is to transfer the polarization state from one side to the other with only a residual constant rotation. \textbf{b} representation of input linear polarizations (in black) and the corresponding output polarization measured after the compensated nanofiber setup (in red). the linearity of the input polarization is maintained. A residual rotation of $11\text{\textdegree} \pm 1\text{\textdegree}$ is observed. \textbf{c} Elliptical polarizations prepared in the entrance of the nanofiber setup and corresponding polarizations measured at the output.}
\label{fig:5}
\end{figure}

The test consists in compensating independently both sides to the nanofiber and testing the transfer of several known polarizations through the overall fiber.
The setup to test the birefringence compensation is presented in figure \ref{fig:5}.(a). 
We use a first Berek compensator (Berek 1) to compensate one side of the fiber.
We then use a second Berek (Berek 2) to independently compensate the birefringence of the other side of the fiber, using light propagating in the opposite direction.

The total optical system, which is the sum of two compensated systems, introduces ideally only a rotation of the input polarization. This rotation is given by the sum of the individual rotations of each fiber portion.

To test this mapping, we prepare different known polarizations and measure the state of polarization at the output of the fiber by means of a polarimeter. 
We prepared a representative panel of polarizations including linearly, elliptically and circularly polarized light using half and quarter wave-plates.
We show in figure \ref{fig:5}.(b),(c) the polarization prepared at the input (black) compared to the polarization measured at the output (red). 
For linear input polarization, the measured output polarization is almost linearly polarized and mainly differ by a rotation of $11.1\text{\textdegree} \pm 0.8\text{\textdegree}$. 
This residual rotation is simply the sum of the individual rotations introduced by each compensated fiber portions: $-64.0\text{\textdegree}$ for the left fiber portion, $+72.9\text{\textdegree}$ for the right fiber portion, resulting in an expected overall rotation of $8.9\text{\textdegree}$, close to the observed value. 
Elliptical and circular polarizations are also well mapped as can be seen in \ref{fig:2}.(c) where the rotation is also noticeable.

The accuracy of the mapping can be quantified in terms of normalized Stokes parameters. 
Compared to the known input normalized Stokes parameters (and correcting for the residual rotation), the output Stokes parameters have a standard deviation of 0.09 corresponding to an accuracy of $95\%$ for the polarization mapping.
Note that this accuracy is given for two compensated systems in series so that the mapping for a single one is expected to be better.

\section{Conclusion}
We demonstrated a new scheme for fiber birefringence compensation in order to achieve full control of the polarization in an optical tapered nanofiber.
The compensation procedure relies on analysing the light scattered by a nanofiber using two imaging systems with $45\text{\textdegree}$-difference camera angle.

We carefully studied the case of light scattering by defects in a nanofiber.
Simulations based on Mie theory indicates that the nanofiber significantly alters the radiation pattern of dipole scatterers embedded in its bulk.
This effect enables to explain the reduced visibility observed for the radiation pattern of a nanofiber scattering its guided light.
Our simulations and observations corroborate the model of point-like scatterer distributed in the nanofiber.
More generally, we show how the far-field radiation properties give information on the microscopic distribution of scatterer in the nanofiber.

Finally, we demonstrated that the compensation procedure enables to transfer polarization state (represented by normalized Stokes parameters) to the nanofiber with an accuracy better than $95\%$.
This work allows for a better (complete) control of nanofiber polarization and opens the way for exotic dipole trapping geometries \cite{REITZ2012} or nanofiber polarization states reading \cite{Joos2018}. 

\section{Acknowledgement}
The authors thanks Clément Sayrin, Baptiste Gouraud, Guillaume Blanquer and Jeremy Raskop for helpful discussions.

\section*{Funding} This work is supported by the Emergence program from Ville de Paris and by PSL Research
University in the framework of the project COSINE.

\appendix
\section{Appendix A: Effect of interference on radiation pattern visibility}
\label{appendix:A}
In the Figure \ref{fig:3} of the main text, we showed that an increasing number of scatterer per unit of optical resolution causes the visibility to increase.
In this section, we want to explain why, based on a detailed description of the spatial dependence of radiation diagrams.

\begin{figure}[htbp]
\scriptsize
\def\svgwidth{\linewidth}
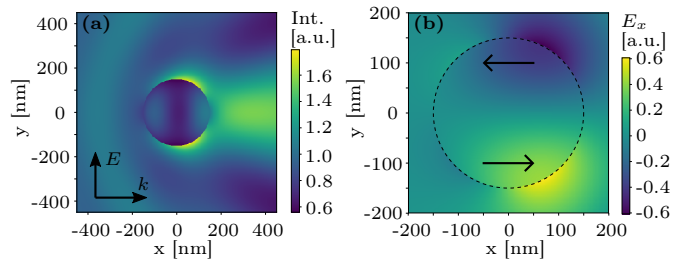
\caption{\textbf{Near field diffraction of a plane wave by a nanofiber}. \textbf{(a)} Intensity distribution for an incident field polarized along y. \textbf{(b)} Plot of the longitudinal electric field component $E_x$ responsible for the radiation of light along the dipole direction when it is located in the nanofiber. The black arrows indicate the direction of the field component illustrating the phase opposition between the lower and upper part of the nanofiber.}
\label{fig:App1}
\end{figure}

Let us consider the diffraction of a plane wave by a nanofiber, the axis of which is supposed to be along $z$.
More specifically, we assume an incident field propagating in the $x\text{-direction}$ and polarized linearly along the $y\text{-direction}$.
We show in figure \ref{fig:App1}.(a) the intensity distribution obtained in the vicinity of the nanofiber for a radiation wavelength of $638\text{ nm}$, a nanofiber diameter of $300\text{ nm}$ and indices of refraction of $1.46$ and $1.00$ for the silica and air respectively.

The strong variation of index of refraction at the nanofiber surface causes the paraxial approximation to break and leads to the emergence of a field component along the direction of propagation.
This longitudinal component of the electric field is shown in figure \ref{fig:App1}.(b).

The mode of propagation considered here is therefore transverse in the far field of the nanofiber (it is a plane wave) but has a longitudinal component in the vicinity of the nanofiber.
A linear dipole located in the nanofiber can hence radiate light along its axis direction.

As can be seen in figure \ref{fig:App1}.(b), the amplitude of the longitudinal component changes sign under reflection at the $xz\text{-plane}$.
This can be understood as a consequence of spin-momentum locking in transversely decaying electric fields \cite{VanMechelen:16}.

Consider now two dipoles contained in the $xy\text{-plane}$ with same orientation (for example horizontal) and symmetrically located with respect to the $x-axis$.
Taken individually, they would each radiate the same amount of light in the $x\text{-direction}$.
But if we consider the coherent radiation of this two dipoles that oscillate in phase, they will interfere destructively in the far field $x\text{-direction}$.
That is, they are not radiating anymore along their dipole axes as they would individually. The visibility of to symmetrical dipoles in the nanofiber is therefore 1.

The visibility of the coherent radiation pattern of a large number of dipoles with same orientation tends to approach unity because each dipole has probably a symmetrical replica to interfere destructively with.

\bibliography{bibliography}

\end{document}